\documentclass{article}
\usepackage{times}  
\usepackage{helvet}  
\usepackage{courier}  
\usepackage{url}  
\usepackage{graphicx}  
\usepackage{amssymb}

\usepackage{amsmath}
\usepackage{wrapfig} 
\usepackage{graphicx}
\usepackage{subfig} 
\usepackage[table]{xcolor}
\usepackage{url}
\usepackage{multirow}
\usepackage{algorithm} 
\usepackage{algorithmic}
\usepackage{amsthm}
\usepackage{url}
\usepackage{times}
\usepackage{helvet}
\usepackage{courier}

\def\tamer{{\textsc{tamer}}}
\def\star{{\textsc{star}}}

\usepackage[font=small]{caption}

\title{Teaching Social Behavior through Human Reinforcement for Ad hoc Teamwork\\ \Large{\textit{The STAR Framework}}}
\author{Shani Alkoby \\
	\texttt{shani.alkoby@gmail.com}
	\and Avilash Rath \\
	\texttt{rathavilash@gmail.com}
	\and Peter Stone \\
	\texttt{pstone@cs.utexas.edu}\\\\
	The University of Texas at Austin, USA
}

\date{}
\begin{document}
	
	\maketitle

\begin{abstract}  
As AI technology continues to develop, more and more agents will become capable of long term autonomy alongside people. Thus, a recent line of research has studied the problem of teaching autonomous agents the concept of ethics and human social norms. Most existing work considers the case of an individual agent attempting to learn a predefined set of rules. In reality however, social norms are not always pre-defined and are very difficult to represent algorithmically. Moreover, the basic idea behind the social norms concept is ensuring that one's actions do not negatively influence others' utilities, which is inherently a multiagent concept. Thus, here we investigate a way to teach agents, as a \emph{team}, how to act according to human social norms. In this research, we introduce the \star\ framework used to teach an \emph{ad hoc team} of agents to act in accordance with human social norms. Using a hybrid team (agents and people), when taking an action considered to be socially unacceptable, the agents receive negative feedback from the human teammate(s) who has(have) an awareness of the team's norms. We view \star\ as an important step towards teaching agents to act more consistently with respect to human morality.
\end{abstract}

\section{Introduction}

Embedding ethics and social norms into AI systems in general, and in the decision making process of autonomous agents in particular, has long been a grand challenge for AI \cite{wallach2008moral}. As a result, many studies focusing on technical approaches for enabling AI systems and autonomous agents to capture the concept of social norms have emerged \cite{yu2018building}. However, despite the abundance of research dealing with the question of how to inject social norms into intelligent agents, there are still three main problems that remain unsolved \cite{rossi2018building}. First, most studies address the problem of a single autonomous AI agent working in isolation. In reality, however, AI agents will increasingly work together in teams that will include humans and other agents as their team members. This discrepancy can lead to many uncertainties and incompatibilities due to policies being tailored to a single agent that may not necessarily be optimal or even relevant in the context of a team. Second, none of the existing studies has offered a way to address the various cultural and temporal dynamics of the broad spectrum of human norms, i.e., the fact that ethics and social norms are not absolute, timeless, universally agreed upon concepts. Third, most approaches use the same formalism for both the function to be maximized and the social boundaries. We note that, even though this can help make the technical calculations easier, due to them being two independent objectives, it may be important to allow for the possibility of goals and norms being represented differently.

In this research we present an approach for teaching autonomous agents the concept of human social norms which provides a solution to the above problems. For this purpose we turn to the ``ad hoc teamwork'' setting in which a team of agents is formed ad hoc, for a particular purpose, and thus the team strategies cannot be developed a priori. Ad hoc teamwork has been studied recently in the AI literature ~\cite{albrecht2013game,AAMAS12-Barrett}. However, to date, no attention has been dedicated to examining whether the methods proposed are \textit{safe} in the sense of preventing the agents from choosing socially unacceptable actions in order to complete their task. We use this scenario in which agents are working together and their actions have mutual influence on one another, to create an online mutual learning process which leads to socially acceptable behavior by the entire team.

In this paper we introduce a novel training paradigm called ``\textbf{S}ocially \textbf{T}raining \textbf{A}gents via \textbf{R}einforcement'' (\textbf{\star}). Using \star\ we study the case of a hybrid team including agents and people. During the cooperation, when taking an action considered to be unacceptable (as opposed to ineffective), the agents receive negative feedback on a dedicated channel for this purpose from the human teammate(s). This will allow  online learning of social codes based on the specific cultural and temporal dynamics relevant to the society the agents are part of. Our method builds upon past work introducing the \tamer\ framework for learning from positive and negative human feedback~\cite{knox2009interactively}.  \tamer\ is based on the assumption that feedback is given to teach the agent how to be more effective. Our work differs by introducing a separate channel by which a person can indicate actions that are unacceptable regardless to how effective they are. Using this social feedback during the learning process, agents are able to develop a set of internal rules such that given a task they will be able to solve it compatibly with the humans' concept of social norms.



\section{Related Work}
This section begins with a survey of the current state of the art in the area of ethics and AI. Next, it discusses reinforcement learning under constraints and its connection to our research. Finally, it provides an overview of the existing literature on ad hoc teamwork.
\subsection{Ethics and AI}
Humans often constrain their decisions according to some exogenous priorities such as morality, ethics, or religion \cite{sen1974choice}. Intelligent agents should be able to do the same. Thus, the AI community is interested in building such smart systems that will be able to restrict their actions by similar principles \cite{arnold2017value,yu2018building,wallach2008moral}. For example, the work of Balakrishnan et al. \cite{balakrishnan2018using,balakrishnan2018incorporating} studies the problem of applying dynamic ethics rules to content recommendation systems. Another example can be the value alignment problem ~\cite{russell2010intelligence,hadfield2017inverse,hadfield2016cooperative,hadfield2018incomplete}. We note that most existing approaches to the value alignment problem assume that misalignment comes from an error in goal specification, inadequate constraints on actions, or lack of human knowledge, whereas we assume the goal specification from the human to be precise and add an additional layer of social norms to the agent's learned behavior. Moreover, most existing work trying to incorporate ethics and social norms into AI agents only considers the case of an individual agent. Those that do consider collective decision making \cite{greene2016embedding}, provide only an initial approach to embedding ethical and social codes into collective decision making.

\subsection{Constrained Reinforcement Learning}
One very relevant line of work is that studying the problem of safe RL (also called RL under constraints) \cite{alshiekh2018safe,serrano2018learning} in which agents try to ensure safety during learning or execution phases of a policy learned using reinforcement learning algorithms. The notion of safety there, however, differs from ours in two ways. First, they represent safety as a set of pre-defined rules (e.g., a finite budget) that can be formulated in a formal way (e.g., using temporal logic) and thus there is no real advantage to the use of human feedback. Second, in their work, an action is considered to be safe only if it does not lead to an undesired state, whereas in our work, an action can lead to a desired state (e.g., having a million dollars) and still be unacceptable (e.g., stealing).  

\subsection{Ad hoc Teamwork}
The design of autonomous agents that can be a part of an ad hoc team is an important open problem in multiagent systems and as such has been widely studied ~\cite{AAMAS12-Barrett,IJCAI11-Zilberstein,macalpine2017evaluating}. Several works addressed this problem by proposing methods which utilize beliefs over a set of hypothetical behaviors for the other agents~\cite{acr2016aij,bs2015,cdzc2014,albrecht2017reasoning}. One crucial issue that is yet to be studied is the question of \textit{how social the agents' actions are on their way to the goal}. This issue is particularly important due to the fact that we are heading toward a future in which agents will be capable of long-term autonomy without direct supervision of humans. Consider for example, future robots which, as part of their daily tasks, may need to wait in lines with people. In this case, knowing the social norms within the community with regards to queuing behaviors could strongly influence, for example, whether a robot should crowd to the front or wait patiently in line. 

\section{Preliminaries}
Our \star\ framework is motivated by \tamer  \cite{knox2009interactively}, an established framework for teaching agents the effectiveness of their actions using human feedback, which models the learning task as a Markov Decision Process. The following subsections provide a formal description of Markov Decision Process and of the \tamer\ framework. 

\subsection{Markov Decision Processes}
As in many decision-making problems, our problem can be represented as a Markov Decision Process (MDP). In MDPs, an agent has a set of possible states denoted by $S$. Additionally, the agent has a set of actions $A$ from which it can choose an action at every time step. Given a state and an action, the probability of transitioning to another state on the next time step is denoted by the transition function $T$, $T: S \times A \times S \rightarrow \mathbb{R}$. A discount factor, $\gamma$, can be used in order to exponentially decrease the value of a future reward. We use $D$ in order to denote the distribution of start states. Finally, a reward function $R$, $R: S \times A \times S \rightarrow \mathbb{R}$, provides the reward received by the agent based on the most recent state, the most recent action, and the next state $s_t$, $a_t$, and $s_{t+1}$. Formally, an MDP can be represented by the tuple ($S$, $A$, $T$, $\gamma$, $D$, $R$).

Many reinforcement learning algorithms \cite{sutton1998rli} seek to learn MDP policies ($\pi : S \rightarrow A$) that maximize return from each state-action pair, where return is equal to $\sum_{t=0}^t \mathbb{E}[\gamma^t R(s_t, a_t)]$.  MDP reward is considered \textit{flawless} since the policies determined by it are optimal (i.e., for each state, choose the action with the highest possible return). However, in many tasks (e.g., chess) the received reward signal is both ``sparse'' and ``delayed''. This is because in many of those tasks, state-action pairs that do not lead to a termination of the task receive zero reward. Thus, the agent must wait until the end of the episode to receive any information from the environment that helps it determine the quality of each state-action pair.

\subsection{The TAMER Framework}
Formally, the \tamer\ framework is an approach to the shaping problem, which is: given a human trainer observing an agent's behavior and delivering evaluative reinforcement signals, how should the agent be designed to make it leverage the human reinforcement signals to learn good behavior? \cite{knox2009interactively}. Using \tamer\ one can replace the sparse and delayed MDP reward signal with a human reward signal.  As opposed to MDP reward signal, when a human trainer observes an agent's behavior, he has a model of the long-term effect of that behavior. Thus, human feedback contains information about whether the targeted behavior is good or bad in the long term. Additionally, since both the time it takes the human trainer to evaluate the targeted behavior and the time it takes to manually deliver the evaluation within a reinforcement signal are relatively small in most cases, human reinforcement is not sparse and is only trivially delayed. On the other hand, due to the fact that humans are boundedly rational \cite{Kahneman2000a,azaria2015strategic,Rabin1998} (and may become tired of giving feedback), their evaluations tend to be imperfect (flawed). Overall, even though the human reward signal is not ``flawless'' as is the MDP reward signal, the fact that the signals provided by humans are not sparse and delayed enables it to learn good behaviors more efficiently ~\cite{knox2009interactively}.

\section{Social Behavior in Ad hoc Teamwork}

The goal of ad hoc teamwork is for an \emph{individual} agent to figure out how best to act in order to contribute to its team's success, \emph{given the behaviors and/or learning strategies of its teammates}. However, to date, team success has not included any notion of social norms.  Thus an ad hoc teammate may take actions that, while in the long-term interest of the team, violate the constraints of social behavior among agents and/or with human teammates. 
Understanding how to inject social norms into the decision making process of the agents is a timely challenge.

We note that the concept of social norms is a notoriously difficult capability to represent algorithmically \cite{cointe2016ethical}. We, therefore, propose that it ought to be taught directly by instructors. Since our objective is to align an agent with subjective human social norms (i.e., we do not rely on there being absolute, universally acceptable norms), humans themselves have the knowledge that can enable the learning process, reducing costly sample complexity. 

In this paper, we address the problem of controlling the agent's social behavior using the framework of ad hoc teamwork and a novel extension of \tamer\, which we introduce in the following section.

\subsection{The STAR Framework}\label{STAR}

Like \tamer, \star\ uses human feedback. \star\ however, does not limit the use of human feedback only to the effectiveness aspect of the action performed, i.e., it has an additional channel by which a person can indicate actions that are unacceptable even when they are technically effective. Based on both the effectiveness signal and the social signal, agents need to find ways to solve the problem that do not violate the social customs. In effect, agents must create a form of ``inner conscience'' helping them to solve a given problem compatibly with humans' social norms.  

\begin{figure} [h]
	\hspace{-10pt}\includegraphics[height=8cm,width=1\linewidth]{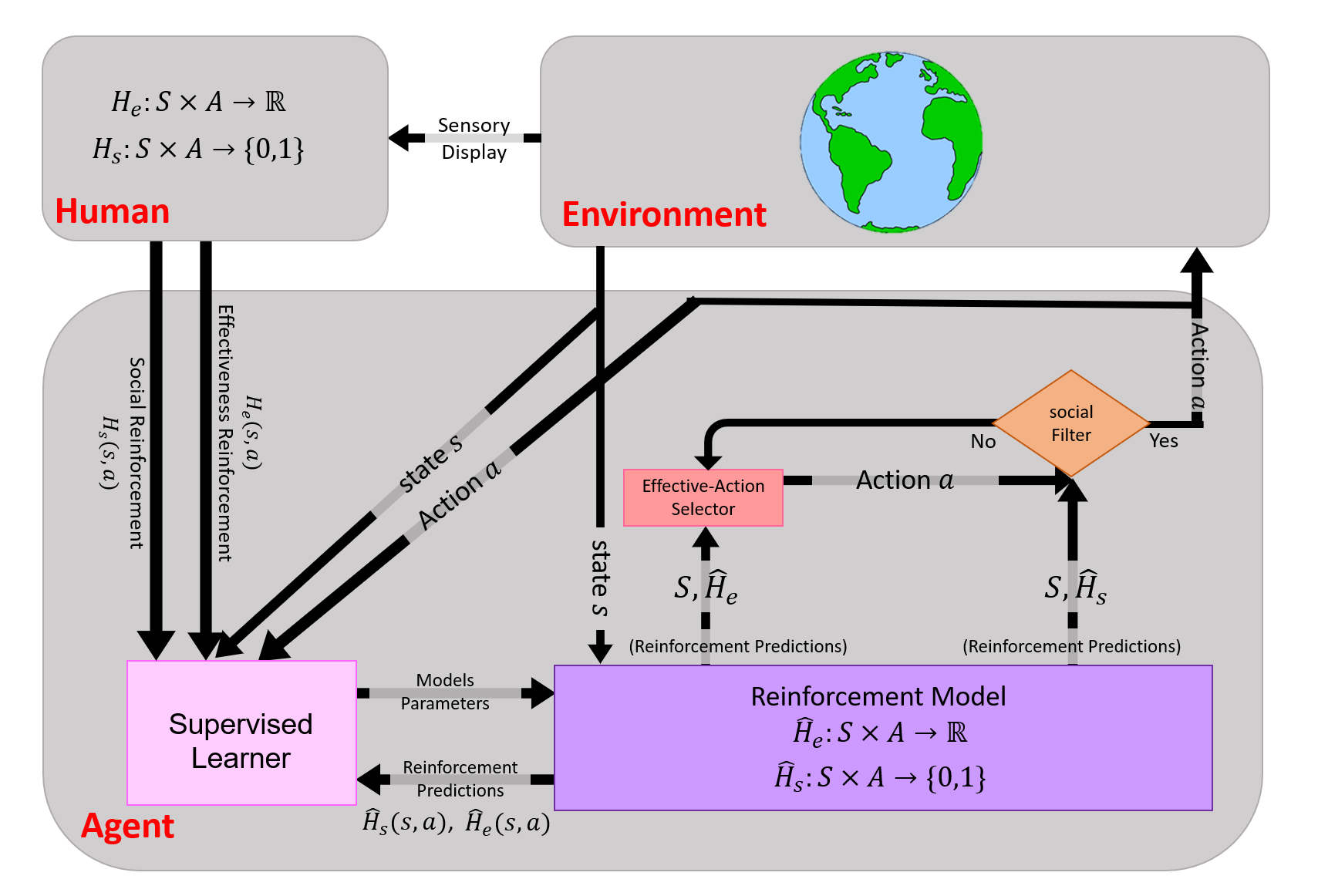}
	\caption{The \star\ framework.}
	\label{fig:STAR_Framework} 
\end{figure}

Figure \ref{fig:STAR_Framework} shows the interaction between a human, the environment, and a \star\ agent within an MDP. In the figure, the human constructs a state $s$ from the environment's display. In addition we assume that the human has both an effectiveness function (i.e., $H_e: S \times A \rightarrow \mathbb{R}$) and a social function (i.e., $H_s: S \times A \rightarrow \{0,1\}$) as internal functions so that given a state $s$, and an action $a$ that the agent has taken, the human is able to provide feedback to the agent that is consistent with them. The agent learns models of these two functions. Using the models, the agent's ``effective-action selector'' chooses an action which is then sent to the ``social filter''. If it passes the filter, in addition to it being performed, the action is also sent to the supervised learner along with the current state as an input. The supervised learner then refines the agent's models based on the information that this action is the most effective action among the permissible actions. Otherwise, the agent chooses a new (predicted to be less effective) action until it finds one that passes the social filter. Finally, we note that in \star\, as in \tamer\, the learning is treated as a supervised learning problem, and does not require value propagation. This is due to the premise that humans provide feedback on the long-term effects of an action - the return, rather than the reward.

\subsection{Assumptions}

\begin{center}
	\begin{figure*}[h]
		\centering
		\includegraphics[height=3cm, width=12cm]{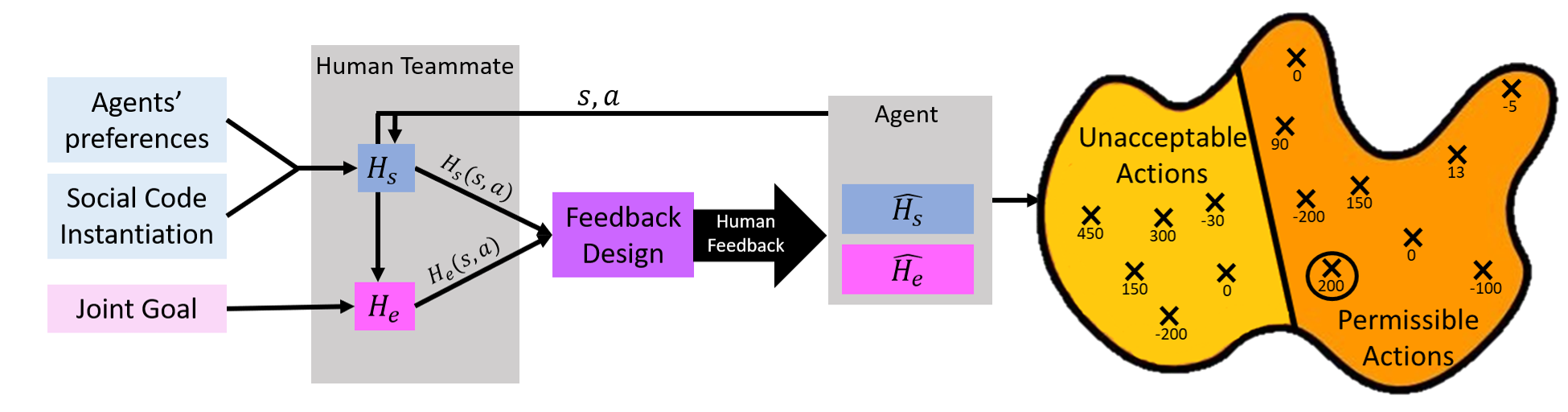}
		\caption{Our experimental evaluation protocol.}
		\label{fig:evaluationsFlow}
	\end{figure*}
\end{center}

We assume that the effectiveness function, $H_e$, is a scalar function, i.e., for each state-action pair, $(s,a)$, given as an input, the function will produce a number representing the effectiveness (value) of taking action $a$ in a state $s$. The social function, $H_s$, in contrast, is a binary function which given a state and action, $(s,a)$, returns $1$ if taking action $a$ in state $s$ is considered to be permissible and $0$ otherwise. We make this simplifying assumption in our initial instantiation of \star\ so as to avoid the necessity of multi-objective optimization. There is ample research on that topic \cite{liu2015multiobjective,roijers2013survey} that can be brought to bear when relaxing this assumption in future work. Additionally we assume that the boundary between the set of permissible actions and the set of unacceptable ones is clear, i.e., given a state, every action is either permissible or unacceptable. Formally: $\forall s\in S, a\in A, H_s(s,a) \in \{0,1\}$. The rightmost part of Figure \ref{fig:evaluationsFlow} illustrates those two assumptions. In the figure, every cross represents a state-action pair (where for all pairs the state is the current state) and the number under each pair represents the action's predicted effectiveness. The actions on the left are considered to be unacceptable while the actions on the right are considered to be permissible. Finally, we assume that for each possible state there exists at least one applicable permissible action, i.e., $\forall s\in S, \exists a\in A \text{ s.t. } H_s(s,a)=1$.

We further assume that social norms are specific to each team of humans. That is, rather than being an absolute, global concept, it is relative to the team on which the agent is participating. However, we do assume that each team has a consistent view of social norms between its members.\footnote{We note that it is straightforward to extend our work to include an overlying concept of global social norms simply by further constraining the set of available actions.} In addition, we note, that the agent's only purpose is to learn the team's social function, and hence it does not judge the team's choices regarding what they consider to be permissible.  Finally, we assume that if it is able to properly learn the team's social function, the agent will always choose a permissible action, i.e., the only cause of an agent choosing an unacceptable action is its lack of ability to properly learn the team's social function. We note that in this research, we are not considering settings in which unacceptable actions can cause real harm. Such actions need to be forbidden entirely, e.g., by law, and are separate from the social norms we aim to teach. We are also aware of the fact that there are cases in which it is challenging for humans to specify the intended social norms, for example the trolley problem \cite{thomson1984trolley}. Our framework is meant for the more common case in which people have a clear understanding of the acceptable social norms, e.g., driving norms in a specific city, where there is generally broad agreement among the people in the society regarding what is permissible and what is not. 

\section{Experimental Evaluation}
\star\ is designed for situations in which a team of agents is working together towards a cooperative goal, while at the same time each maintains individual preferences.  For example, a construction crew may have the cooperative goal of building a house, with each person preferring different types of houses (floor plans, decors, etc.), and each preferring not to do more than his or her share of the work. To represent such tasks in an easily controllable experimental setting, we introduce a new domain, \textit{Multiagent Tetris} (MaT), with the following essential properties: (i) a team of agents works together towards the joint goal of clearing as many rows as possible as a team; (ii) each agent has individual \textit{preferences} for intermediate states along the way towards goal states; (iii) a variety of \textit{social codes} can be defined indicating the team's joint attitude towards how much each other's preferences ought to be taken into account when selecting actions (these codes can range from considering only one's own preferences, to making sure never to violate any teammate's preferences); (iv) when provided with the details of the joint goal, the different agents' preferences, and the particular instantiation of team social code, a person can provide feedback to the agent both on the effectiveness of any given action towards the joint goal and whether the action is consistent with the team's social code. Our experimental evaluation protocol is illustrated in Figure \ref{fig:evaluationsFlow}. 

In the following subsections, we provide a detailed description of the MaT domain and indicate how each of these properties is met.  In Section \ref{Sec:Results}, we evaluate different designs of the \star\ framework that ablate various aspects of the user's feedback signal.

\subsection{The MaT Domain}

\begin{wrapfigure}{r}{0.15\textwidth}
	\includegraphics[width=0.15\textwidth]{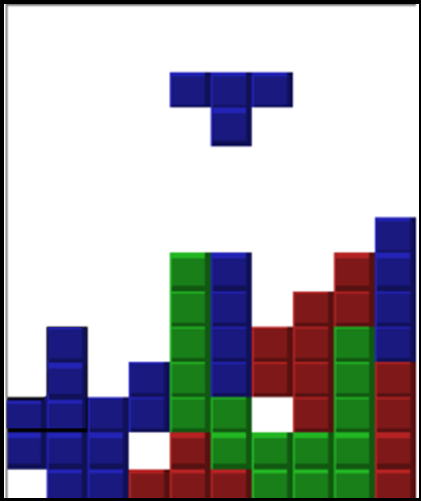}
	\caption{A screen shot of the MaT domain.}
	\label{fig:screenshot}
\end{wrapfigure}

We use Tetris as the basis for our experimental domain due to the fact that it has already been shown that \tamer\ performs well in it \cite{knox2009interactively}. In MaT, a number of agents take turns playing in the same game. They use round-robin scheduling such that in each agent's time slice that agent is the only one playing. We define an agent's time slice to be the amount of time in which the agent controls $n\in \mathbb{N}$ blocks. The agent's slice ends when it puts down its last of the $n$ blocks. As a result, an agent is able to control its blocks from beginning to end and is able to place them exactly where it intends to. 

Additionally, blocks in MaT are of different colors (as elaborated below). The agents share the joint goal of clearing as many rows as possible, i.e., the score of each agent is a function of the overall number of rows cleared by the entire team. Since clearing a row requires a sequence of actions to be performed, the team has to collaborate in order to achieve this goal. Figure \ref{fig:screenshot} depicts a screen shot of the MaT game.

\subsection{Agents' Preferences}\label{subsec:preferences}
As mentioned above, each agent has individual \textit{preferences} for intermediate states along the way towards goal states. In practice this means that each agent is differently influenced by different adjacent pairs of colors placed on the board. Thus, in contrast to the score received from accomplishing the joint goal of clearing a row, which influences all agents in the same way, each intermediate state of the board has a different influence on each of the agents. 

\subsection{Social Code Instantiations}\label{subsec:instantiation}
A team's social code indicates to what degree each agent should take into account its own and the other agents' preferences when selecting actions.  For example, in one team, it may be considered permissible to always take the action which leads to the most rows being cleared, whereas in another team it may never be considered permissible to take an action that goes against another teammate's preferences. 

To numerically represent each agent's affinity for an intermediate board state, we sum up the number of adjacent pairs this agent finds to be good and subtract the number of adjacent pairs it considers to be bad. Each adjacent pair which is \textit{strongly} preferred by the agent is added twice to the overall sum, and those that are \textit{strongly} disliked by the agent are subtracted twice from the overall sum. There may also be adjacent pairs that the agent is indifferent to, and naturally those will not influence its affinity for the board. For example, if one agent likes red next to blue, \textit{strongly} dislike blue next to green, and is indifferent to all other combinations, its affinity for the board state in Figure \ref{fig:screenshot} is $-10$ (since there are $6$ red-blue pairs and $8$ blue-green pairs that are being subtracted twice).

In this paper we consider two possible instantiations of the social code concept, \textit{global} and \textit{simple}.\footnote{We note that both the \textit{global} and \textit{simple} social codes are common in many real life societies and thus were chosen to be discussed in the paper. However, other social codes are also possible and can be learned using our framework.} If using the global social code, one should consider the influence of its action over the whole team's resulting board affinity. Given an action $a_i$ performed by an agent, we will define $\Delta_{a_i}$ to be the difference between the sum of all teammates' board affinities after performing $a_i$ and before performing it. Thus, given an action $a_i$, it is considered to be unacceptable only if $\Delta_{a_i}$ is negative. We note that according to this definition it does not matter if some of its teammates lose from the agent's action, as long as $\Delta_{a_i}$ is positive, (i.e., the team as a whole has benefit from the action), this is a permissible action. This type of social code focuses on maximizing social welfare. If using the simple social code however, any action that causes any damage to any other agent is considered to be unacceptable.

\subsection{Feedback Design}\label{subsec:feedbackDesign}
During execution, one or more human teammates are able to give feedback to the agents regarding the action most recently selected. Since we assume that all teammates share the same social norms, in all of our experiments, we include just a single human teammate giving advice to all the agents. This person is told both the rules of the game and which instantiation of social code to follow, and is instructed to provide separate feedback on an action's effectiveness and the degree to which it aligns with the social code.  

\begin{figure}[h]
	\centering
	\includegraphics[height=5cm,width=12cm]{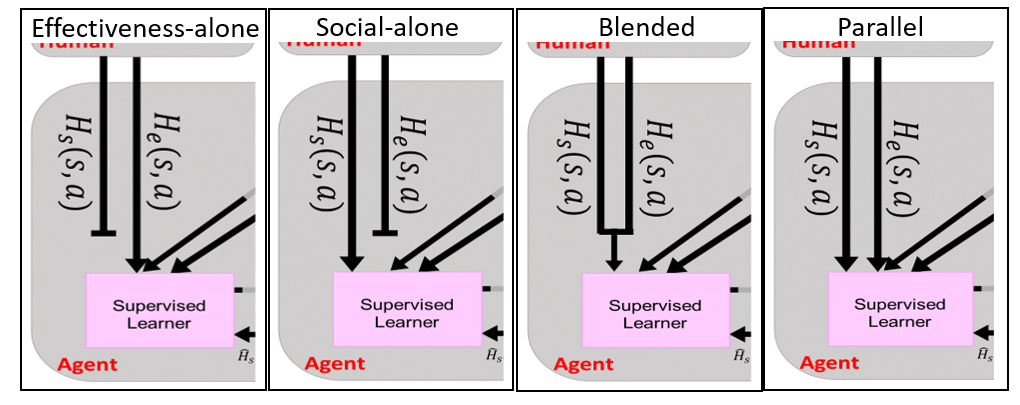}
	\caption{Four \star\ designs with varied feedback received by agents, illustrated as modifications to the left side of Figure \ref{fig:STAR_Framework}.}
	\label{fig:4designe}
\end{figure}


We implemented four \star\ designs as illustrated in Figure \ref{fig:4designe}. In all cases, the human received the same instructions and the same interface for providing feedback, which included separate buttons to indicate whether an action is effective vs. permissible. However, we varied the information that the agents received as follows. Our original \star\ system provides the feedback from two separate parallel channels, each specifically designed for providing human feedback only regarding the effectiveness/social aspect of its actions. This design is called the \textit{parallel} feedback design. Additionally, as baselines, we considered two single feedback designs in which the agents take into account only one aspect of the feedback provided. In the \textit{effectiveness-alone} feedback design, the agent receives feedback only regarding the effectiveness of its performed action whereas in the \textit{social-alone} feedback design, it receives feedback only regarding the social aspect. Each one of those designs is used to check how an agent is able to learn one concept, effectiveness or human social norms, with no other distractions. A higher level of baseline is a feedback design that does take into account both the effectiveness and the permissibility of a given action but eliminates the separation between the two. This design is called the \textit{blended} feedback design. In the \textit{blended} feedback design the agent receives positive feedback on the effectiveness channel if the action is effective or permissible and negative feedback if it is ineffective or unacceptable,  i.e., this design creates an identity between the two feedback types. We note that the \textit{effectiveness-alone} design is identical to the original \tamer\ framework design for the cases where the feedback provider only considers the effectiveness aspect of the actions performed. If however, the feedback provider also considers the social aspect while providing the feedback, it is actually the \textit{blended} feedback design that is identical to the design of the original \tamer\ framework. By comparing the agent's learned behavior when using these different designs we will be able to characterize the influence each feedback channel has on the performance of the agents. 

\section{Results}\label{Sec:Results}

\begin{wrapfigure}{r}{0.2\textwidth}
	\includegraphics[width=0.2\textwidth]{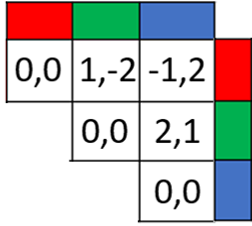}
	\caption{Agents' preferences toward different adjacent pairs of blocks}
	\label{fig:matrix}
\end{wrapfigure}

Though in principle, MaT can be scaled up to include many agents, for the sake of simplicity, in all experiments reported in this paper we use a team composed of two agents playing and one human as a feedback provider. Additionally, we used three possible colors (red, green and blue) for the Tetris blocks. Figure \ref{fig:matrix} illustrates the different agents' preferences. Of the nine possible adjacent pairs of color blocks, both agents were indifferent towards adjacent pairs of blocks of the same color (i.e., red-red, green-green, and blue-blue). The first agent preferred red-green, \textit{strongly} preferred green-blue, and disliked red-blue. The second agent preferred green-blue, \textit{strongly} preferred red-blue, and \textit{strongly} disliked red-green. Both agents receive feedback from their human teammate. We evaluated the two instantiations of social code described in Section \ref{subsec:instantiation}, both with the full \star\ system and with 3 ablations that remove or modify aspects of the feedback provided by the human teammate as described in Section \ref{subsec:feedbackDesign}. A game ends when the board is completely full, i.e., there is no more space for new blocks to enter. 

\begin{figure}[ht]
	\hspace{-10pt}
	\includegraphics[height = 9cm, width=13cm]{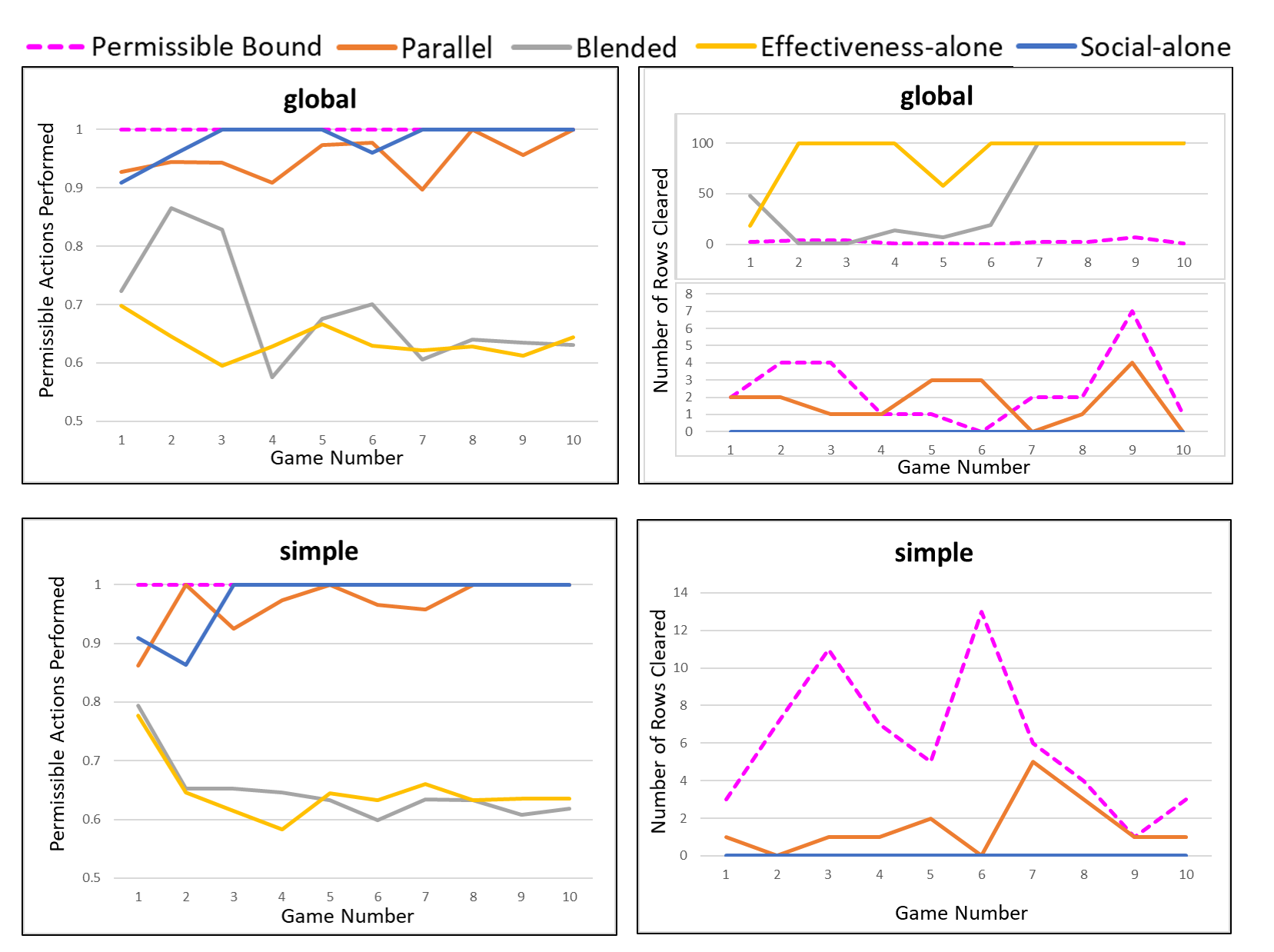}
	\caption{Agents' performance for each of the four possible designs using each of the social code instantiations for all ten games.}
	\label{fig:graphs}
\end{figure}

It has been shown in previous work (\cite{knox2009interactively}) that when receiving effectiveness feedback from a human,  agents learn the effective policy rapidly. Thus, in our experiments, we measure the performance only until the tenth game. For the results to be comparable we eliminated randomness in block ordering and tested the different designs on an identical set of ten games. Additionally, for consistency in the human feedback, we created a human proxy that provides the feedback to the agents based on a pre-defined policy taught by a human. Hence, the policy used as the basis for the human proxy is suboptimal (as would be any real human's policy). In principle it should be possible to teach a more effective policy with an optimal proxy. However, since the main focus of our work is on the concept of permissibility, this suboptimality does not impact our results. Using the described setting, we hypothesized that the \textit{parallel} feedback design leads to the best team performance within the constraints of the team's social code i.e., using each of the baseline feedback designs will lead to more unacceptable actions being selected, less effective behavior, or both.

Figure \ref{fig:graphs} illustrates the agents' performance for each one of the possible designs under each of the social code instantiations for all ten games. Since our aim is to teach a team of agents to act according to the human social norms, and since compliance with social norms can only cause a decrease in the performance, it might be the case where the team does learn, however does not perform well due to the social constrains. Therefore, to be able to properly evaluate the different designs' performance, we calculated an upper bound for the number of rows that can be cleared under only permissible actions given the social code. The upper left graph depicts the percentages of permissible actions out of all the actions performed in each game for the \textit{global} social code instantiation and the lower left graph depicts the percentages of permissible actions for the \textit{simple} one. The two right graphs depicts the effectiveness performance of the team, i.e., the number of rows cleared in each game, for the \textit{global} and \textit{simple} social code respectively. We note that for the \textit{simple} social code the number of rows cleared in both \textit{effectiveness-alone} and the \textit{blended} designs was much higher than the number of rows cleared in all other designs and was not on the same scale. Thus, to maintain the graph's readability we do not include them in the figure.


As expected, regardless of the social code we are using, when the agents receive only social feedback, the number of rows cleared is the lowest. The percentage of the permissible actions used however, is the highest. On the contrary, if receiving effectiveness feedback only, the team is able to clear the highest number of rows per game, but use a relatively high number of unacceptable actions in order to do so. Under the \textit{blended} feedback design, the team's performance is quite good and is somewhat close to their performance under the \textit{effectiveness-alone} design. The number of permissible actions under this design is also very low, and in most cases is similar to the number of permissible actions used under the \textit{effectiveness-alone} design. The similarity between the two designs' curves shows that even if the agents receive feedback both regarding the effectiveness and the social aspect of their actions, they do not learn well when the feedback is provided without the proper context. Hence, given that our goal is to teach a team what are human social norms, the \textit{blended} design is ineffective for this purpose. Finally, if using the original \star\ framework, i.e., the \textit{parallel} design, we can see a learning pattern for both objectives. Note, that in the last three games, after the team has learned, the parallel curve gets very close to the upper bound and even starts following its shape. This shows that, using the \textit{parallel} design, the team is able to  learn both the effectiveness and the social functions such that their performance is very close to the best performance possible given a specific social code. These results shows that the use in the \star\ framework enables us to teach a team of agents to act in accordance with human social norms without damaging their performance given the permissibility constraints.

\section{Discussion and Future Work}
In the last decade, many works have studied the problem of teaching autonomous agents how to behave socially for preventing them from acting in ways which are contrary to the good of society. Those studies mainly concentrated on how to teach a single agent the concept of ethics and social norms. In this paper we introduce the \star\ framework meant to teach ad hoc team how to work together under the constraints of human social norms.  \star\ addresses the social aspect of autonomous agents' behavior as part of a long-term autonomy process and enables agents to learn, from experience, what is considered to be a permissible or an unacceptable behavior using the interactions it has with the people around it. Testing several designs for providing an agent with human feedback for two common instantiations of social codes, we confirmed our hypothesis that the best way to teach an agent how to be both social and effective is by using our full \star\ framework, i.e., using two parallel feedback channels, each devoted to providing only effectiveness/social feedback. We note that in our framework one can trivially replace the method by which the agents learn the target function (e.g., combine RL with human feedback \cite{knox2010combining}), thus  opening the door for hybrid approaches to embedding ethics and social norms into AI systems.  

There are many future research directions opened up by this research. Our ultimate goal is to teach agents the concept of human morality for creating a safe agent behavior. However, for this purpose, some of the assumptions made in this work will have to be relaxed. For example, one of the main assumptions in this research is the dichotomy property of $H_s$, i.e., an action can be either permissible or unacceptable. However, when it comes to morality this is not the case due to the existence of a moral hierarchy. For instance, it is \textit{more immoral} to kill someone than to steal a dollar. Thus, a natural next step is relaxing the dichotomy assumption and, using a multiobjective optimization, to explore a case in which an agent needs, in addition to learning what is considered to be permissible and what is not, to learn the full, complex concept of morality.

\newpage
\bibliographystyle{plain}

\end{document}